\documentclass[conference]{IEEEtran}
\IEEEoverridecommandlockouts
\ifCLASSINFOpdf
\usepackage[pdftex]{graphicx}
\else
 \usepackage[dvips]{graphicx}

\fi

\usepackage[cmex10]{amsmath}

\usepackage{cases}
\usepackage{amsfonts}	
\newcommand{\R}{\mathbb{R}}
\newcommand{\prob}{\mathbf{P}}
\newcommand{\Ex}{\mathbf{E}}

\newcommand{\pcero}{\mathbf{P}^0}
\newcommand{\pout}{\mathbf{P}_{out}}
\newcommand{\poutcf}{\mathbf{P}_{out,CF}}
\newcommand{\poutdf}{\mathbf{P}_{out,DF}}
\newcommand{\poutpp}{\mathbf{P}_{out,PP}}

\begin{document}
%
\title{Cooperative Strategies for  Interference-Limited Wireless Networks}

\author{\IEEEauthorblockN{Andr\'{e}s Altieri\thanks{This work was partially supported by the Peruilh grant of the University of Buenos Aires.}, Leonardo Rey Vega, Cecilia G. Galarza}
\IEEEauthorblockA{
School of Engineering \\
University of Buenos Aires and CONICET\\
Paseo Col\'{o}n 850, Buenos Aires, Argentina\\
Email: aaltieri, lrey, cgalar@fi.uba.ar}
\and
\IEEEauthorblockN{Pablo Piantanida}
\IEEEauthorblockA{Department of Telecommunications\\
SUPELEC\\
91192 Gif-sur-Yvette, France\\
Email: pablo.piantanida@supelec.fr}
}

\maketitle

\begin{abstract}
Consider the communication of a single-user aided by a nearby relay involved in a large wireless network where the nodes form an homogeneous Poisson point process. Since this network is interference-limited the asymptotic error probability is bounded from above by the outage probability experienced by the user. We investigate the outage behavior for the well-known cooperative schemes, namely, decode-and-forward (DF) and compress-and-forward (CF). In this setting, the outage events are induced by both fading and the spatial proximity of neighbor nodes who generate the strongest interference and hence the worst communication case. Upper and lower bounds on the  asymptotic error probability which are  tight in some cases are derived. It is shown that there exists a clear trade off between the network density and the benefits of user cooperation. These results are useful to evaluate performances and to optimize relaying schemes in the context of large wireless networks. 
\end{abstract}

\IEEEpeerreviewmaketitle

\section{Introduction}
Spatial models for large wireless networks where nodes are not assumed to have much interference management have received much attention in recent years. An elegant and comprehensive analysis framework for these networks has been developed through the use of stochastic geometry and random graphs (see \cite{HABDF2009} and references therein). Performance metrics such as transport capacity \cite{GK2000}, outage probability (OP) and transmission capacity \cite{WYAV2005, WAJ2010} have been introduced to evaluate performance and scaling laws of the mentioned networks. 

In terms of traditional Shannon theory, these novel metrics essentially rely on a fundamental result in information theory known as Feinstein's Lemma \cite{Feinstein-1954}. This applies to any homogeneous network, providing an upper bound on the error probability for the communication of each user. Hence the probability of ``unsuccessful communication", as stated in \cite{BBM2006}, turns to be an upper bound on the asymptotic error probability for the communication between any pair of nodes in the network. Similarly, from \cite{verdu-te-sum-han-1994} it can be shown that such probability also gives a lower bound on the error probability. It clearly states that under homogeneous processes the error probability of users behaves as the OP, where for spatial networks user mobility and fading induce these outage events. 

The main limiting factor of self-organizing and mobile ad-hoc networks is the average distance between sources and their destinations. Such networks are not assumed to need much interference management which results in a constant capacity per link. However this assumption is not quite realistic; while coordination can be light, all real networks need some control if they wish to stay operational for some time. 
The simplest scenario of cooperation consists of a single-relay which helps communication between the transmitter (source) and the receiver (destination). Cover and El Gamal \cite{CG1979} developed the main strategies, namely, Decode-and-Forward (DF) and Compress-and-Forward (CF), and showed an upper bound referred to as the cut-set bound. Capacity is only known for some special cases (see  \cite{Kramer2005} and references therein). 
 
In this work, we investigate how performance, measured in terms of OP, behaves when a transmitter in a large wireless network is aided by a nearby relay. We model the network as an independently marked Poisson point process and assume that the network is interference-limited, that is, the performance is limited by the signal-to-interference ratio (SIR). The signal attenuation is assumed to occur both through path-loss and fading, and the encoder uses either the DF or CF coding schemes. The SIR-based outage events are determined from the information-theoretic achievable rates \cite{CG1979}. Users send Gaussian symbols at a fixed rate and hence an outage is declared whenever node distribution or fading causes this rate to be higher than the achievable rate \cite{CG1979}. Closed-form expressions for the OP of both protocols under Rayleigh fading are derived. For the DF scheme, the OP can be calculated exactly in terms of the Laplace transform of the process \cite{BB2010} while in the CF case, we managed to develop an upper bound which is tight for small network densities. For the DF scheme, optimal correlation between source and relay signals is discussed. Finally the performance of both protocols is compared against direct transmission \cite{BBM2006} and the cut-set bound, which provides a lower bound on the OP of any coding scheme.


\section{System Model}

Our planar network is modeled as an independently marked Poisson point process:
\begin{equation}
\tilde{\Phi} = \left\{ (x_i,h_{x_ir},h_{x_id})\right\}. \label{eq:PPP}
\end{equation}
\begin{itemize}
\item The set of transmitters constitutes an homogeneous Poisson point process $\Phi = \left\{ x_i \right\}$ of intensity $\lambda$.
\item We assume that a transmitter node located at the origin attempts to communicate to a destination at location $d=(D,0)$ with the aid of a relay located at $r$. The relay position $r$, known by the transmitter, can be parameterized without loss of generality as:
\begin{equation}
r = k D \left( \cos \theta, \sin\theta\right),
\end{equation}	
with $\theta \in [0;2\pi)$ and $k>0$.
\item All users transmit with constant unit power. The power received at $y$ by a transmitter at $x$ is $|h_{xy}|^2 l(|x-y|)$ where:
\begin{itemize}
\item $l(|x-y|)$ is the spherically symmetric path-loss between $x$ and $y$. For our numerical results we shall work with the usual simplified path-loss function:
\begin{equation}
l(x,y) = |x-y|^{-\alpha}
\end{equation}
with $\alpha >2$. For shortness we shall sometimes write $l_{xy} := l(|x-y|)$.
\item $|h_{xy}|^2$ is the power fading coefficient associated with the channel between points $x$ and $y$. We consider Rayleigh fading, i.e. the power fading coefficients are independent identically distributed exponential random variables with unit mean.
\end{itemize}
\item The marks $h_{x_ir}$ and $h_{x_id}$ model the fading coefficient between each transmitting node in the network and the nodes relay and destination corresponding to the transmitter located at the origin, respectively. In addition we include another fading coefficient $h_{rd}$ with the same distribution as $h_{x_ir}$ and $h_{x_id}$, independent of $\tilde{\Phi}$, which models the fading between the relay and destination corresponding to the transmitter at the origin.
\end{itemize}
In dealing with the relay channel we have to work simultaneously with the interference produced by all the sources at two different points in space. Since these paths have different fading coefficients, we define two separate interference fields:
\begin{gather}
I_d = \sum_{x \in \Phi} |h_{xd}|^2 l(|x-d|), \\
I_r = \sum_{x \in \Phi} |h_{xr}|^2 l(|x-r|).
\end{gather}
We also define analogous fields $I^0_d$ and $I^0_r$ whose expressions are the same as $I_d$ and $I_r$ respectively, which do not include the node at the origin in the summation.

\section{Decode-and-Forward (DF) Scheme}

The DF strategy allows the relay to decode the messages sent by the source, re-encode them, and forward them to the destination. In this setting, the encoder cannot optimize the rate $R$ since it is unaware of the instantaneous interference and fading coefficients involved in the different channels. Thus the error probability  is dominated by the OP. 

 \subsection{Conditions for outage}
Consider the achievable rate for the DF scheme \cite{CG1979} and specialize it to our channel. Given a rate $R$ we shall consider the outage event: $\mathcal{A}_{DF}\cup \mathcal{B}_{DF}$, where:
\begin{gather*}
\mathcal{A}_{DF}=\left\{ \frac{|h_{sr}|^2 l_{sr}\left(1-|\rho|^2\right)}{I^0_r}<T \right\}, \\
\mathcal{B}_{DF}\hspace{-0.3mm}= \hspace{-0.3mm}\left\{ \frac{|h_{sd}|^2 l_{sd}+|h_{rd}|^2 l_{rd} + 2\sqrt{l_{sd} l_{rd}} \Re(\rho h_{sd}h_{rd}^*)}{I^0_d} <T\right\} \hspace{-1mm}.
\end{gather*}
The event $\mathcal{A}_{DF}$ means that the relay is in outage and the event $\mathcal{B}_{DF}$ means that the destination is in outage. $T= 2^R-1$ and $\rho$ is the complex correlation coefficient between the symbols transmitted by the source and the relay. $(\cdot)^*$ denots complex conjugation and $\Re(\cdot)$ denotes the real part of a complex number.
\subsection{Outage probability}
To determine the outage probability $\poutdf$ for the DF protocol we need to evaluate the following expression:
\begin{multline}
\poutdf = 1 - \pcero\left\{\frac{|h_{sr}|^2 l_{sr}\left(1-|\rho|^2\right)}{I^0_r}\geq T \right.,\\
 \left. \frac{|h_{sd}|^2 l_{sd}+|h_{rd}|^2 l_{rd}+ 2\sqrt{l_{sd}l_{rd}} \Re(\rho h_{sd}h_{rd}^*)}{I^0_d}\geq T\right\},
\end{multline}
where $\pcero$ is the Palm probability of the stationary marked point process having a point at the origin. 
Using Slivnyak's Theorem \cite{BB2010,DVJ1998} this can be rewritten as:
\begin{multline}
\poutdf = 1-\prob\left(|\hat{h}_{sr}|^2 \geq \frac{TI_r}{l_{sr}\left(1-|\rho|^2\right)} \right., \\
\left. \frac{|\hat{h}_{sd}|^2 l_{sd}+|\hat{h}_{rd}|^2 l_{rd}+ 2\sqrt{l_{sd} l_{rd}} \Re(\rho \hat{h}_{sd} \hat{h}_{rd}^*)}{I_d}\geq T\right), \label{eq:poutDF0}
\end{multline}
where $\hat{h}_{sr}$, $\hat{h}_{sd}$ and $\hat{h}_{rd}$ have the same distribution as $h_{sr}$, $h_{sd}$ and $h_{rd}$ but are independent of $\tilde{\Phi}$. Defining:
$V:= |\hat{h}_{sd}|^2 l_{sd}+|\hat{h}_{rd}|^2 l_{rd}+ 2\sqrt{l_{sd} l_{rd}} \Re(\rho \hat{h}_{sd} \hat{h}_{rd}^*)$, 
and considering that $\hat{h}_{sr}$ and $V$ are independent of each other and of $\tilde{\Phi}$ we have that:
\begin{equation}
\poutdf = 1 - \Ex_{\tilde{\Phi}} \left[ e^{-\frac{TI_r}{l_{sr}\left(1-|\rho|^2\right)}}F_V(TI_d)\right] \label{eq:PoutDF1}
\end{equation}
where $F_V(\cdot)$ is the complementary distribution function of $V$. For Rayleigh fading it is straightforward to show that:
\begin{numcases}{F_V(s) = }
\frac{\mu_2 e^{-s / \mu_2}-\mu_1 e^{- s / \mu_1}}{\mu_2-\mu_1} & $\mu_1 \neq \mu_2$ 	 \label{eq:mu1mu2} \\
(1+s/\mu_1) e^{-s/\mu_1} & $\mu_1 = \mu_2 $
\end{numcases}
where:
\begin{gather}
\mu_{1} = \frac{(l_{sd} + l_{rd}) - \left( (l_{sd} - l_{rd})^2 + 4 l_{sd} l_{rd} |\rho|^2 \right)^{1/2}}{2}  \label{eq:mu1}\\
\mu_{2} = \frac{(l_{sd} + l_{rd}) + \left( (l_{sd} - l_{rd})^2 + 4 l_{sd} l_{rd} |\rho|^2 \right)^{1/2}}{2}. \label{eq:mu2}
\end{gather}
Since $\mu_1= \mu_2$ only if $\rho = 0$ and $k = 2 \cos(\theta)$ we focus our attention to the case $\mu_1 \neq \mu_2$. Replacing (\ref{eq:mu1mu2}) in (\ref{eq:PoutDF1}) and rearranging the terms gives:
\begin{multline}
\poutdf = 1 - \frac{\mu_2}{\mu_2-\mu_1} \mathcal{L}_{I_d,I_r} (T/\mu_2,T/\mu_3)+ \\
\frac{\mu_1}{\mu_2-\mu_1} \mathcal{L}_{I_d,I_r} (T/\mu_1,T/\mu_3),\label{eq:PoutDF2}
\end{multline}
where $\mu_3 = l_{sr}\left(1-|\rho|^2\right)$ and $\mathcal{L}_{I_d,I_r}\left( \cdot, \cdot\right)$ is the joint Laplace transform of the interference at the relay and at the destination \cite{BB2010}. For the simplified path loss function it is straightforward to show that:
\begin{equation}
\mathcal{L}_{I_d,I_r}(\omega_1, \omega_2) = e^{-\lambda \left(C (\omega_1^{2/\alpha}+ \omega_2^{2/\alpha}) + f(\omega_1, \omega_2) \right)}, \label{eq:Lap2D}
\end{equation}
where:
\begin{gather}
C = \frac{2\pi\Gamma\left(\frac{2}{\alpha}\right) \Gamma\left(1-\frac{2}{\alpha}\right)}{\alpha} \label{eq:C} \\
f(\omega_1 , \omega_2) = \int_{\R^2} \frac{\omega_1 \omega_2}{(\omega_1+ |x-d|^\alpha) (\omega_2+ |x-r|^\alpha)} dx
\end{gather}
and $\Gamma(z) = \int_0^\infty t^{z-1} e^{-t} dt$ is the Gamma function. An interesting remark is that although we have considered unit mean, the DF outage probability is in fact independent of the fading mean, as in the point to point case \cite{BBM2006}.

For the DF protocol we need to choose the appropriate value of the correlation coefficient $\rho$ which minimizes the outage probability. Notice that the outage probability is dependent only on the absolute value of $\rho$ and not on its phase. Moreover, the outage probability is actually a monotone increasing function of $|\rho|$ in the interval $[0,1]$. The optimal choice in terms of outage probability is therefore $\rho = 0$. For space limitations, we skip the proof of this statement.

\section{Compress-and-Forward (CF) Scheme}

The CF strategy allows the relay to compress the received signal and forward it to the destination without decoding the messages. Wyner-Ziv coding is used for optimal compression. Optimization of the rate $R$ is not possible since the source is unaware of the instantaneous interference and fading. Thus the error probability is dominated by the OP. 

\subsection{Conditions for outage}
Consider the achievable rate for the CF scheme \cite{CG1979} and  specialize it to our channel. Given a rate $R=\log(1+T)$ we shall consider the outage event:
 $\mathcal{A}_{CF}\cup  \mathcal{B}_{CF}$, where:
\begin{gather}
\mathcal{A}_{CF}= \left\{ \frac{|h_{sr}|^2l_{sr}}{I^0_r+W_c} + \frac{|h_{sd}|^2 l_{sd}}{I^0_d} < T \right\},  \label{eq:ET}\\
\mathcal{B}_{CF}= \left\{ W_c < \frac{I^0_d l_{sr}|h_{sr}|^2+I^0_rl_{sd} |h_{sd}|^2 + I^0_r I^0_d}{l_{rd} |h_{rd}|^2} \right\}.
\end{gather}
The event $\mathcal{B}_{CF}$ means that relay-destination link cannot sustain the rate needed to transmit the compressed version of the received signal, while $W_c$ denotes the power of the noise added to compress the channel output of the relay and send it to the destination.

\subsection{Outage probability}
To compute the outage probability $\poutcf$ we have to evaluate the reduced Palm probability of $\mathcal{A}_{CF}\cup  \mathcal{B}_{CF}$. By using a similar argument as the one used for the DF case, we obtain the expression:
\begin{multline}
\poutcf = \prob \left( \frac{|\hat{h}_{sr}|^2l_{sr}}{I_r+W_c} + \frac{|\hat{h}_{sd}|^2 l_{sd}}{I_d} < T \right. \cup \\
\left. W_c < \frac{I_d l_{sr}|\hat{h}_{sr}|^2+I_rl_{sd} |\hat{h}_{sd}|^2 + I_r I_d}{l_{rd} |\hat{h}_{rd}|^2} \right)
\end{multline}
where once more $\hat{h}_{sr}$, $\hat{h}_{sd}$ and $\hat{h}_{rd}$ have the same distribution as $h_{sr}$, $h_{sd}$ and $h_{rd}$ respectively but are independent of $\tilde{\Phi}$. The outage event for this case is more involved than the one for the DF case, so instead of finding the actual outage probabilities we provide an upper bound which is valid when the node distribution is sparse ($\lambda$ small). 
In terms of $\mathcal{A}_{CF}$ and  $\mathcal{B}_{CF}$ we can write:
\begin{equation}
\poutcf = \prob(\mathcal{A}_{CF}) + \prob(\bar{\mathcal{A}}_{CF} \cap \mathcal{B}_{CF}) \label{eq:PoutCF1}
\end{equation}
where $\bar{(\cdot)}$ denotes set complement. The probability $\prob(\mathcal{A}_{CF})$ can be upper bounded to a desired degree of accuracy. A simple bound would be to take:
\begin{equation}
\prob(\mathcal{A}_{CF}) \leq \prob \left( \frac{|\hat{h}_{sr}|^2l_{sr}}{I_r+W_c} < T, \frac{|\hat{h}_{sd}|^2 l_{sd}}{I_d} <T \right). \label{eq:simplebound}
\end{equation}
In this bound we cover the set $\mathcal{A}_{CF}$ by a square of side $T$. Such bound can be tightened as desired by partitioning the set $\mathcal{A}_{CF}$ further and covering it by disjoint rectangles, yielding:
\begin{multline*}
\prob(\mathcal{A}_{CF}) \leq 
\prob\left(\frac{|\hat{h}_{sr}|^2 l_{sr}}{I_r+W_c} < T \right) - \\
 \sum_{n=0}^{N-1} \hspace{-.7mm} \prob \hspace{-.7mm} \left(\frac{n}{N}T \hspace{-.1mm} \leq  \hspace{-.1mm} \frac{|\hat{h}_{sr}|^2 l_{sr}}{I_r+W_c} \hspace{-.1mm} < \hspace{-.1mm} \frac{n+1}{N} T \right. \hspace{-0.6mm} ,\hspace{-0.6mm}
\left. \frac{|\hat{h}_{sd}|^2 l_{sd}}{I_d} \hspace{-.1mm} \geq \hspace{-.1mm}\frac{N-n}{N} T\right) \hspace{-0.9mm},
\end{multline*}
which can be written in a straightforward manner in terms of the Laplace transform of the interference. Taking $N=1$ gives (\ref{eq:simplebound}), which is useful theoretical purposes, while higher values of $N$ provide tighter upper bounds for numerical results. 

The second event in (\ref{eq:PoutCF1}) can be upper bounded as follows:
\begin{multline}
\hspace{-3mm} \prob(\bar{\mathcal{A}}_{CF} \cap \mathcal{B}_{CF}) \leq  
 \prob \left( \frac{ |\hat{h}_{sr}|^2 l_{sr}  I_d + |\hat{h}_{sd}|^2 l_{sd} I_r}{W_c l_{rd} |\hat{h}_{rd}|^2} > \frac{T}{1+T} \right) \\
 \hspace{-1.5mm} = 1 \hspace{-0.3mm} - \hspace{-0.3mm} \Ex \hspace{-0.4mm}\left[ \mathcal{L}_{I_d,I_r}\hspace{-0.9mm} \left( \hspace{-0.4mm}\frac{(1+T)l_{sr} |\hat{h}_{sr}|^2}{T W_c l_{rd}},\frac{(1+T)l_{sd} |\hat{h}_{sd}|^2}{T W_c l_{rd}} \right) \hspace{-0.5mm} \right] \hspace{-3mm} \label{eq:PoutCF2}
\end{multline}
where the expectation is taken with respect to $|\hat{h}_{sr}|^2$ and $|\hat{h}_{sd}|^2$. By using this bound we avoid working with the product of the interference at the relay and the destination, which complicates the evaluation of $\prob(\mathcal{B}_{CF})$ significantly. In addition, notice that the product term decreases as $\lambda$ decreases, indicating this bound may be tight for small values of $\lambda$. Discarding the term $f(\omega_1, \omega_2)$ in (\ref{eq:Lap2D}) we further bound: 
\begin{multline}
\prob(\bar{\mathcal{A}}_{CF} \cap \mathcal{B}_{CF}) \leq  1 - \Ex\left[ \mathcal{L}_{I_d}	\left(\frac{(1+T) l_{sr} |\hat{h}_{sr}|^2}{T W_c l_{rd}}\right)\right]\\ \times \Ex\left[ \mathcal{L}_{I_r}\left(\frac{(1+T) l_{sd} |\hat{h}_{sd}|^2}{T W_c l_{rd}}\right)\right]. \label{eq:PoutCF3}
\end{multline}
Since the Laplace transform of the interference is known in closed form for the simplified path loss function (take $\omega_1 = 0$ in (\ref{eq:Lap2D})), these expectations can be calculated numerically without difficulty. Extensive simulation have shown that (\ref{eq:PoutCF3}) is practically the same as (\ref{eq:PoutCF2}). This is possibly due to the fact that the values of $|\hat{h}_{sr}|^2$ and $|\hat{h}_{sd}|^2$ for which $f(x,y)$ is large with respect to the other terms in (\ref{eq:Lap2D}) have a small probability of occurrence, and therefore they have little impact on the expected value of the Laplace transform.
\begin{figure}[!t]
\centering
\includegraphics[width=0.72\columnwidth,keepaspectratio,trim= 53mm 84mm 53mm 85mm,clip]{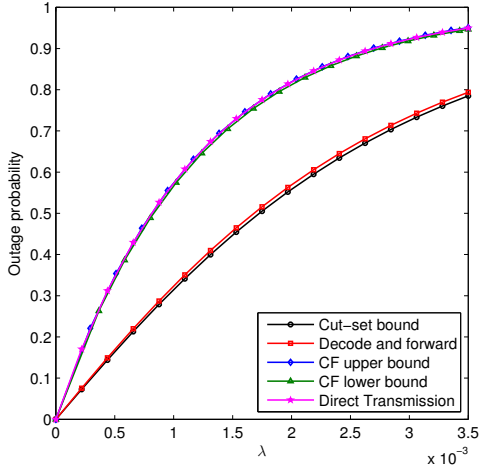} 
\caption{Outage behavior when the relay is close to the destination. $D=10$, $T=3$, $\alpha = 4$, linear geometry ($\theta = 0$) with $k=0.2$. $\rho = 0$ for DF.}
\label{fig:outagek2}
\end{figure}

To evaluate the performance of the upper bound on $\poutdf$ we can develop a lower bound on the outage probability by simply lower bounding $\prob(\mathcal{A}_{CF})$. The bound is obtained in a similar fashion as the upper bound, lower bounding $\mathcal{A}_{CF}$ by disjoint rectangles.

\section{Numerical Results}
The numerical results were obtained with the simplified path loss function only. In order to establish the outage performance of the analyzed protocols we consider the direct transmission case whose outage probability $\poutpp$ is known to be \cite{BBM2006}:
\begin{equation}
\poutpp = 1-e^{\lambda T^{2/\alpha} D^2 C}, \label{eq:Poutpp}
\end{equation}
where $C$ is given by (\ref{eq:C}). 
\subsection{Lower bound on the error probability}
We use the cut-set upper bound \cite{CG1979} to establish the overall performance of the protocols, i.e., a lower bound on the outage probability of any arbitrary coding scheme. Using Slivnyak's theorem, we write this bound in terms of the SIR as:
\begin{multline*}
\max \left\{ \prob \left((1-|\rho|^2)\left(\frac{|\hat{h}_{sr}|^2 l_{sr}}{I_r}+ \frac{|\hat{h}_{sd}|^2 l_{sd}}{I_d}\right) < T \right) \right. ,\\
\left. \prob \left( \frac{|\hat{h}_{sd}|^2 l_{sd}+|\hat{h}_{rd}|^2 l_{rd} + 2\sqrt{l_{sd} l_{rd}} \Re(\rho \hat{h}_{sd}\hat{h}_{rd}^*)}{I_d} < T\right)\right\}.
\end{multline*}
The event in the first term is the same as (\ref{eq:ET}) taking $W_c =0$ and replacing $T$ by $T/(1-|\rho|^2)$. Therefore, it can be bounded to any degree of accuracy in the same way. The second event is similar to the second event in (\ref{eq:poutDF0}) and can be evaluated in a similar fashion as the DF outage probability. 
\begin{figure}[!t]
\centering
\includegraphics[width=0.72\columnwidth,keepaspectratio,trim= 53mm 84mm 53mm 85mm,clip]{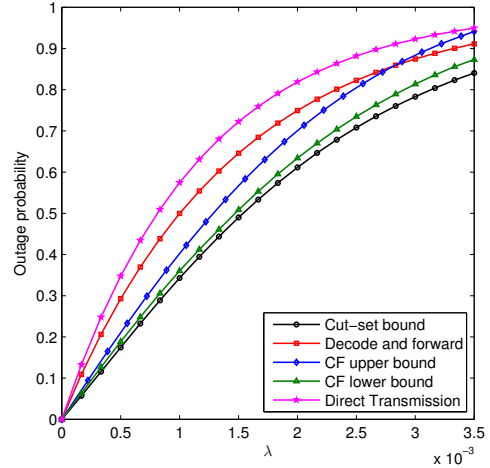}
\caption{Outage behavior when the relay is close to the destination. $D=10$, $T=3$, $\alpha = 4$, linear geometry ($\theta = 0$) with $k=0.9$. $\rho = 0$ for DF.}
\label{fig:outagek9}
\end{figure}
\subsection{Comments on plots}
In all simulations we set $D=10$ and a path loss exponent $\alpha = 4$. For the DF protocol we always take $\rho = 0$ which minimizes the OP, while for the CF protocol and the cut-set bound we numerically optimize for each value of $\lambda$ the values of $W_c$ and $\rho$ respectively. To compare the outage probabilities for different network densities we present a case in which the relay is close to the source and another in which the relay is close to the destination. In both cases we take $T=3$ and for simplicity we consider that the relay, source and destination are aligned, that is, $\theta = 0$. For CF we plot both the upper bound for $\poutcf$ and the lower bound for $\prob(\mathcal{A}_{CF})$ to asses how close the upper bound is to the true CF performance.

In Fig. \ref{fig:outagek2} we consider the case in which the relay is closer to the source by taking $k = 0.2$. First notice that in this case the outage probabilities for DF are very close to cut-set bounds which confirm that DF is optimal when the relay is close to the source. DF performance is closer to the cut-set bound as $k$ becomes smaller, and grows further apart as $k$ grows. On the other hand, for CF notice that the upper and lower bounds are almost the same, which indicates that the upper bound is tight. Since in this case the relay-destination link is as bad as the source-destination link, $W_c$ is high, which means that the relay is transmitting very little information, and hence the overall performance of the channel is very close to direct transmission.

In Fig. \ref{fig:outagek9} we consider the case in which the relay is closer to the destination by taking $k=0.9$. In this case, DF performance is significantly degraded with respect to the cut-set bound while for small values of $\lambda$ CF performance is much closer to it. Notice that the upper and lower bounds for CF are close, specially for small $\lambda$, which means that the upper bound is close to actual CF performance. On the other hand, when the node density grows CF starts to lose performance. This is due to the fact that increasing node density reduces the quality of the relay-destination link.

\begin{figure}[!t]
\centering
\includegraphics[width=0.72\columnwidth,keepaspectratio,trim= 53mm 84mm 53mm 85mm,clip]{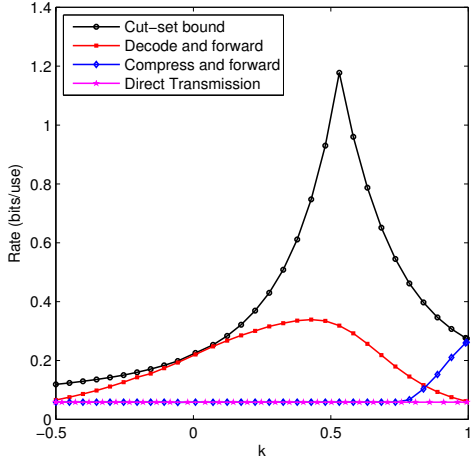} 
\caption{Maximum achievable rates for linear geometry for $\lambda = 10^{-5}$ and $\pout = 10^{-3}$. $D=10$, $T=3$, $\alpha = 4$, $\theta = 0$, $\rho = 0$ for DF. }
\label{fig:rates}
\end{figure}

Now we analyze the maximum rates for each protocol given an outage probability. We fix $\pout = 10^{-3}$ and $\lambda = 10^{-5}$ and evaluate the maximum values of $T$ that can be achieved by each protocol for different values of $k$ and considering linear geometry ($\theta = 0$). In Figure \ref{fig:rates} we can see the maximum rates $R_{\max} = \log_2(1+T_{\max})$ that can be achieved by each protocol in comparison to the cut-set bound and the direct transmission case. We observe that DF is preferred when the relay is closer to the source and CF is preferred when the relay is very close to the destination.

\begin{figure}[!t]
\centering
\includegraphics[width=0.72\columnwidth,keepaspectratio,trim= 53mm 84mm 53mm 85mm,clip]{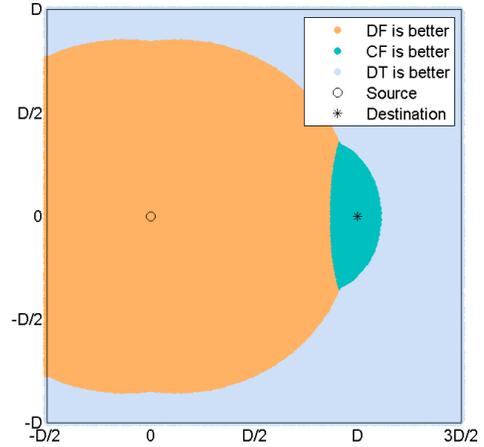} 
\caption{Regions in which DF, CF or direct transmission are preferable for $\lambda = 10^{-4}$. $D=10$, $T=3$, $\alpha = 4$, $\rho = 0$ for DF.}
\label{fig:region}
\end{figure}

Finally in Fig. \ref{fig:region} we plot the regions in space in which DF, CF or direct transmission are preferred over the others. To do this we compare the outage probabilities of each protocol for $\lambda = 10^{-4}$ and chose the one who has the smallest value. We can see that DF is better than the other protocols for a large area around the source. When the relay is very close to the destination, CF dominates over the other transmission strategies. It is interesting to observe that optimal region for DF is significantly larger than the optimal one corresponding to CF.

\section{Summary and Discussions}

In this paper we studied the performance, measured in terms of OP, of a user in a large wireless network when transmitting with the aid of a nearby relay. We studied both DF and CF, the main communication strategies for the relay channel as defined in \cite{CG1979}.
The performances of DF and CF strategies are directly related to the distance between the relay and the destination. In other words, DF scheme performs better than CF when the relay is near to the source while CF scheme is preferable when the relay is near to the destination. 

As future work, it would be of interest to extend the results in the present work to the scenario where the relay position is random and the source is oblivious to the relying strategy and to the presence of the relay (as investigated in \cite{Behboodi}).

\bibliographystyle{IEEEtran}
\bibliography{IEEEabrv,Isit2011}

\begin{thebibliography}{10}
\providecommand{\url}[1]{#1}
\csname url@samestyle\endcsname
\providecommand{\newblock}{\relax}
\providecommand{\bibinfo}[2]{#2}
\providecommand{\BIBentrySTDinterwordspacing}{\spaceskip=0pt\relax}
\providecommand{\BIBentryALTinterwordstretchfactor}{4}
\providecommand{\BIBentryALTinterwordspacing}{\spaceskip=\fontdimen2\font plus
\BIBentryALTinterwordstretchfactor\fontdimen3\font minus
  \fontdimen4\font\relax}
\providecommand{\BIBforeignlanguage}[2]{{%
\expandafter\ifx\csname l@#1\endcsname\relax
\typeout{** WARNING: IEEEtran.bst: No hyphenation pattern has been}%
\typeout{** loaded for the language `#1'. Using the pattern for}%
\typeout{** the default language instead.}%
\else
\language=\csname l@#1\endcsname
\fi
#2}}
\providecommand{\BIBdecl}{\relax}
\BIBdecl

\bibitem{HABDF2009}
M.~Haenggi, J.~Andrews, F.~Baccelli, O.~Dousse, and M.~Franceschetti,
  ``Stochastic geometry and random graphs for the analysis and design of
  wireless networks,'' \emph{{IEEE} J. Sel. Areas Commun.}, vol.~27, no.~7, pp.
  1029 --1046, Sep. 2009.

\bibitem{GK2000}
P.~Gupta and P.~Kumar, ``The capacity of wireless networks,'' \emph{{IEEE}
  Trans. Inf. Theory}, vol.~46, no.~2, pp. 388 --404, Mar. 2000.

\bibitem{WYAV2005}
S.~Weber, X.~Yang, J.~Andrews, and G.~de~Veciana, ``Transmission capacity of
  wireless ad hoc networks with outage constraints,'' \emph{{IEEE} Trans. Inf.
  Theory}, vol.~51, no.~12, pp. 4091 --4102, Dec. 2005.

\bibitem{WAJ2010}
S.~Weber, J.~Andrews, and N.~Jindal, ``An overview of the transmission capacity
  of wireless networks,'' \emph{{IEEE} Trans. Commun.}, vol.~58, no.~12, pp.
  3593 --3604, Dec. 2010.

\bibitem{Feinstein-1954}
A.~Feinstein, ``A new basic theorem of information theory,'' \emph{Information
  Theory, IRE Professional Group on}, vol.~4, no.~4, pp. 2 --22, Sep. 1954.

\bibitem{BBM2006}
F.~Baccelli, B.~Blaszczyszyn, and P.~Muhlethaler, ``An aloha protocol for
  multihop mobile wireless networks,'' \emph{{IEEE} Trans. Inf. Theory},
  vol.~52, no.~2, pp. 421 -- 436, Feb. 2006.

\bibitem{verdu-te-sum-han-1994}
S.~Verd\'{u} and T.~S. Han, ``A general formula for channel capacity,''
  \emph{{IEEE} Trans. Inf. Theory}, vol.~40, pp. 1147--1157, 1994.

\bibitem{CG1979}
T.~Cover and A.~Gamal, ``Capacity theorems for the relay channel,''
  \emph{{IEEE} Trans. Inf. Theory}, vol.~25, no.~5, pp. 572 -- 584, Sep. 1979.

\bibitem{Kramer2005}
G.~Kramer, M.~Gastpar, and P.~Gupta, ``Cooperative strategies and capacity
  theorems for relay networks,'' \emph{{IEEE} Trans. Inf. Theory}, vol.~51,
  no.~9, pp. 3037--3063, Sept. 2005.

\bibitem{BB2010}
F.~{B}accelli and B.~{B}laszczyszyn, \emph{{S}tochastic {G}eometry and
  {W}ireless {N}etworks}.\hskip 1em plus 0.5em minus 0.4em\relax {N}o{W}
  {P}ublishers, 2009.

\bibitem{DVJ1998}
D.~J. Daley and D.~Vere-Jones, \emph{An introduction to the theory of point
  processes. {V}ol. {II}}, 2nd~ed.\hskip 1em plus 0.5em minus 0.4em\relax New
  York: Springer, 2008.

\bibitem{Behboodi}
A.~Behboodi and P.~Piantanida, ``Broadcasting over the relay channel with
  oblivious cooperative strategy,'' in \emph{Communication, Control, and
  Computing (Allerton), 2010 48th Annual Allerton Conference on}, Sep. 2010,
  pp. 1098 --1103.

\end{thebibliography}

\end{document}